\documentclass[11pt,a4paper]{article}
\usepackage{amsmath,amssymb}
\usepackage{epsfig,graphicx}

\begin{document}

\title{\bf Density matrix of the Universe reloaded: origin of inflation
and cosmological acceleration}
\author{A.~O.~Barvinsky$^a$\footnote{{\bf e-mail}: barvin@td.lpi.ru},
C.~Deffayet$^b$\footnote{{\bf e-mail}: deffayet@iap.fr},
A.~Yu.~Kamenshchik$^{c,d}$\footnote{{\bf e-mail}:
Alexander.Kamenshchik@bo.infn.it}
\\
$^a$ \small{\em Theory Department, Lebedev Physics Institute} \\
\small{\em Leninski Prospect 53, Moscow 119991, Russia}\\
$^b$ \small{\em APC UMR 7164 (CNRS, Univ. Paris 7, CEA, Obs. de Paris)} \\
\small{\em 12 rue Alice Domon et L\'eonie Duquet, 75205 Paris Cedex
13, France}\\
$^c$\small{\em Dipartimento di Fisica and INFN}\\
\small{\em via Irnerio 46, 40126 Bologna, Italy}\\
$^d$\small{\em L.D.Landau Institute for Theoretical Physcis}\\
\small{\em Kosygin str. 2, 119334 Moscow, Russia}}
\date{}
\maketitle

\begin{abstract}
We present an overview of a recently suggested new model of quantum
initial conditions for the Universe in the form of a cosmological
density matrix. This density matrix originally suggested in the
Euclidean quantum gravity framework turns out to describe the
microcanonical ensemble in the Lorentzian quantum gravity of
spatially closed cosmological models. This ensemble represents an
equipartition in the physical phase space of the theory (sum over
everything), but in terms of the observable spacetime geometry it is
peaked about a set of cosmologies limited to a bounded range of
the cosmological constant. This suggests a mechanism to
constrain the landscape of string vacua and a possible solution
to the dark energy problem in the form of the quasi-equilibrium
decay of the microcanonical state of the Universe. The effective
Friedmann equation governing this decay incorporates the effect of
the conformal anomaly of quantum fields and features  a new
mechanism for a cosmological acceleration stage -- big boost
scenario. We also briefly discuss the relation between our model,
the AdS/CFT correspondence and RS and DGP braneworlds.
\end{abstract}

\section{Introduction}

It is widely recognized that Euclidean quantum gravity (EQG) is a
lame duck in modern particle physics and cosmology. After its summit
in early and late eighties (in the form of the cosmological
wavefunction proposals \cite{HH,tunnel} and baby universes boom
\cite{baby}) the interest in this theory gradually declined,
especially, in cosmological context, where the problem of quantum
initial conditions was superseded by the concept of stochastic
inflation \cite{stochastic}. EQG could not stand the burden of
indefiniteness of the Euclidean gravitational action \cite{GHP} and
the cosmology debate of the tunneling vs no-boundary proposals
\cite{debate}.

Thus, a recently suggested EQG density matrix of the Universe
\cite{slih} is hardly believed to be a viable candidate for the
initial state of the Universe, even though it avoids the infrared
catastrophe of small cosmological constant $\Lambda$, generates an
ensemble of quasi-thermal universes in the limited range of
$\Lambda$, and suggests a strong selection mechanism for the
landscape of string vacua \cite{slih,lcb}. Here we want to give a
brief overview of these results and also justify them by deriving
from first principles of Lorentzian quantum gravity applied to a
microcanonical ensemble of closed cosmological models \cite{why}. In
view of the peculiarities of spatially closed cosmology this
ensemble describes ultimate (unit weight) equipartition in the
physical phase space of the theory. This can be interpreted as a sum
over Everything, thus emphasizing a distinguished role of this
candidate for the initial state of the Universe.

We analyze the cosmological evolution in this model with the initial
conditions set by the instantons of \cite{slih,lcb}. In particular,
we derive the modified Friedmann equation incorporating the effect
of the conformal anomaly at late radiation and matter domination
stages \cite{BDK}. This equation shows that the vacuum (Casimir)
part of the energy density is "degravitated" via the effect of the
conformal anomaly -- the Casimir energy does not weigh. Moreover,
together with the recovery of the general relativistic behavior,
this equation can feature a stage of cosmological acceleration
followed by what we call a {\em big boost} singularity \cite{BDK}.
At this singularity the scale factor acceleration grows in finite
cosmic time up to infinity with a finite limiting value of the
Hubble factor, when the Universe again enters a quantum phase
demanding for its description an UV completion of the low-energy
semiclassical theory. Then we discuss the hierarchy problem in this
scenario which necessarily arises when trying to embrace within one
model both the inflationary and acceleration (dark energy) stages of the
cosmological evolution. The attempt to solve this problem via the
(string-inspired) concept of evolving extra dimensions brings us to
the AdS/CFT and braneworld setups \cite{AdS/CFT,Randall:1999vf,Gubser,HHR},
including the Randall-Sundrum and DGP models tightly linked by
duality relations to our anomaly driven cosmology.

\section{Euclidean quantum gravity density matrix}

\begin{figure}[h]
\centerline{\epsfxsize 4.4cm \epsfbox{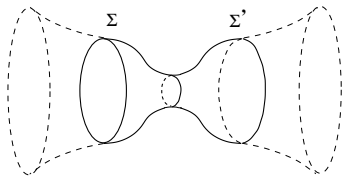}} \caption{\small The
picture of Euclidean spacetime underlying the EQG density matrix,
whose two arguments are associated with the surfaces $\Sigma$ and
$\Sigma'$. Dashed lines depict the Lorentzian signature spacetime
nucleating at $\Sigma$ and $\Sigma'$. \label{Fig.1}}
\end{figure}

A density matrix $\rho(\,q,q')$ in Euclidean quantum gravity
\cite{Page} is related to a spacetime having two disjoint boundaries
$\Sigma$ and $\Sigma'$ associated with its two entries $q$ and $q'$
(collecting both gravity and matter observables), see
Fig.\ref{Fig.1}. The metric and matter configuration on this
spacetime $[\,g,\phi\,]$ interpolates between $q$ and $q'$, thus
establishing mixing correlations.

\begin{figure}[h]
\centerline{\epsfxsize 4.3cm \epsfbox{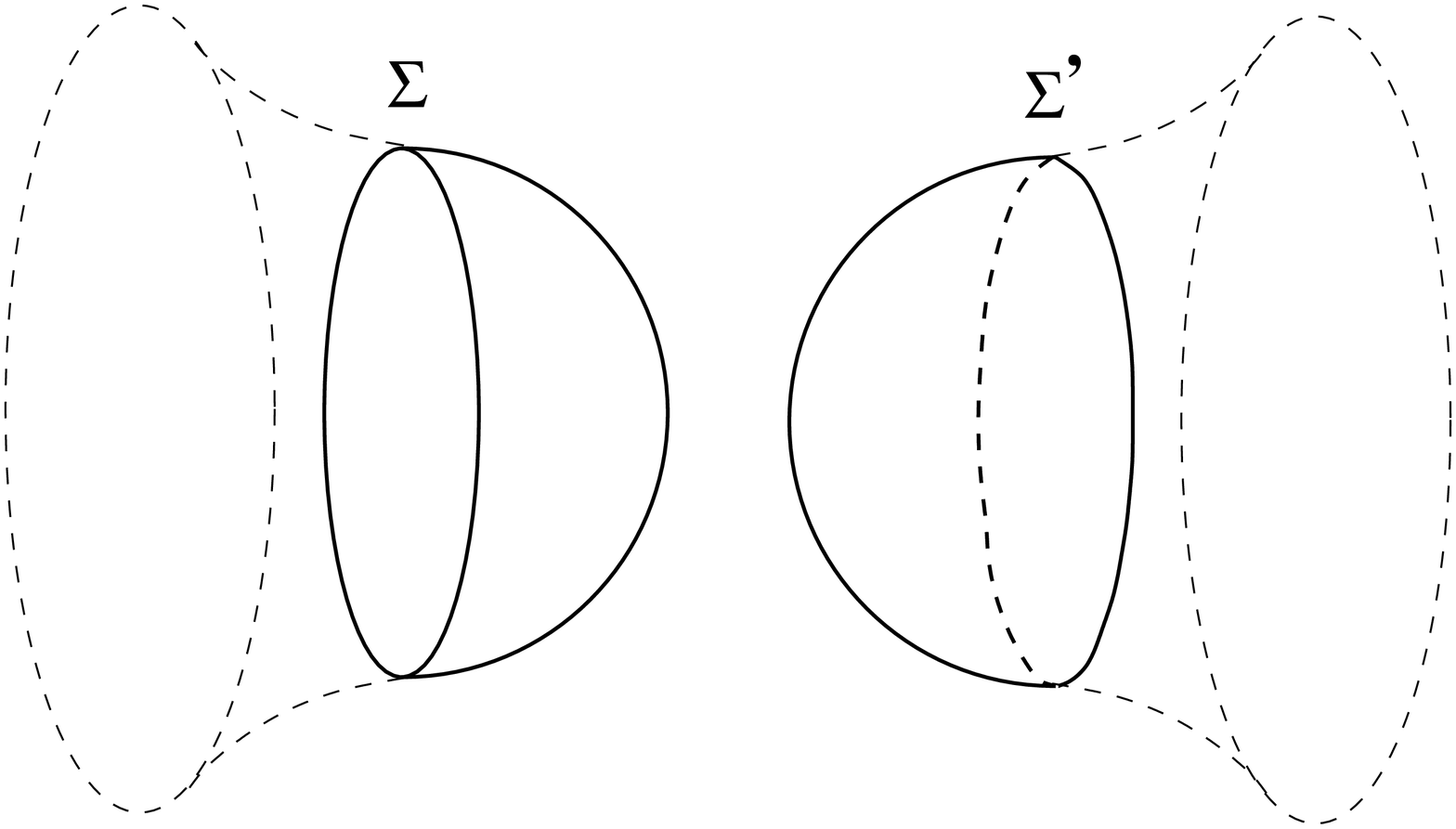}} \caption{\small
Density matrix of the pure Hartle-Hawking state represented by the
union of two instantons of vacuum nature. \label{Fig.2}}
\end{figure}

This obviously differs from the pure Hartle-Hawking state
$|\Psi_{HH}\rangle$ which can also be formulated in terms of a
special density matrix $\hat\rho_{HH}$. For the latter the spacetime
bridge between $\Sigma$ and $\Sigma'$ is broken, so that the
spacetime is a union of two disjoint hemispheres which smoothly
close up at their poles (Fig.\ref{Fig.2}) --- a picture illustrating
the factorization of
$\hat\rho_{HH}=|\Psi_{HH}\rangle\langle\Psi_{HH}|$.

Analogously to the prescription for the Hartle-Hawking state
\cite{HH}, the EQG density matrix can be defined by the path
integral \cite{slih,lcb} over gravitational $g$ and matter $\phi$
fields on the spacetime of the above type interpolating between the
observables $q$ and $q'$ respectively at $\Sigma$ and $\Sigma'$,
    \begin{eqnarray}
    \rho(\,q,q'\,)=
    e^\varGamma
    \int
    D[\,g,\phi\,]\,
    \exp\big(-S_E[\,g,\phi\,]\big),             \label{rho0}
    \end{eqnarray}
where $S_E[\,g,\phi\,]$ is the classical Euclidean action of the
system. In view of the density matrix normalization ${\rm
tr}\hat\rho=1$ the corresponding statistical sum $\exp(-\varGamma)$
is given by a similar path integral,
    \begin{eqnarray}
    e^{-\varGamma}=\!\!\int\limits_{\,\,\rm periodic}
    \!\!\!\! D[\,g,\phi\,]\,
    \exp\big(-S_E[\,g,\phi\,]\big),   \label{statsum}
    \end{eqnarray}
over periodic fields on the toroidal spacetime with identified
boundaries $\Sigma$ and $\Sigma'$.

For a closed cosmology with the $S^3$-topology of spatial sections
this statistical sum can be represented by the path integral over
the periodic scale factor $a(\tau)$ and lapse function $N(\tau)$ of
the minisuperspace metric
    \begin{eqnarray}
    ds^2 = N^2(\tau)\,d\tau^2+a^2(\tau)\,d^2\Omega^{(3)} \label{FRW}
    \end{eqnarray}
on the toroidal $S^1\times S^3$ spacetime \cite{slih,lcb}
    \begin{eqnarray}
    &&e^{-\varGamma}=\!\!\int\limits_{\,\,\rm periodic}
    \!\!\!\! D[\,a,N\,]\;
    e^{-\varGamma_E[\,a,\,N\,]},   \label{1}\\
    &&e^{-\varGamma_E[\,a,\,N]}
    =\!\!\int\limits_{\,\,\rm periodic}
    \!\!\!\! D\varPhi(x)\,
    e^{-S_E[\,a,\,N;\,\varPhi(x)\,]}.             \label{2}
    \end{eqnarray}
Here $\varGamma_E[\,a,\,N]$ is the Euclidean effective action of all
inhomogeneous ``matter" fields which include also metric
perturbations on minisuperspace background
$\varPhi(x)=(\phi(x),\psi(x),A_\mu(x)$, $h_{\mu\nu}(x),...)$,
$S_E[a,N;\varPhi(x)]\equiv S_E[\,g,\phi\,]$ is the original
classical action of the theory under the decomposition of the full
configuration space into the minisuperspace and perturbations
sectors,
    \begin{eqnarray}
    [\,g,\phi\,]=
    [\,a(\tau),N(\tau);\,\varPhi(x)\,],    \label{decomposition}
    \end{eqnarray}
and the integration also runs over periodic fields $\varPhi(x)$.

Under the assumption that the system is dominated by free matter
fields conformally coupled to gravity this action is exactly
calculable by the conformal transformation taking the metric
(\ref{FRW}) into the static Einstein metric with $a={\rm const}$
\cite{slih}. In units of the Planck mass $m_P=(3\pi/4G)^{1/2}$ the
action reads
    \begin{eqnarray}
    &&\varGamma_E[\,a,N\,]=m_P^2\int d\tau\,N \left\{-aa'^2
    -a+ \frac\Lambda3 a^3
    +B\!\left(\frac{a'^2}{a}
    -\frac{a'^4}{6 a}\right)
    +\frac{B}{2a}\,\right\}
    +F(\eta), \label{FrieEu}
    \end{eqnarray}
where
    \begin{equation}
    F(\eta)=\pm\sum_{\omega}\ln\big(1\mp
    e^{-\omega\eta}\big),\,\,\,\,\,
    \eta=\int d\tau N/a,                 \label{effaction}
    \end{equation}
and $a'\equiv da/Nd\tau$. The first three terms in curly brackets
represent the classical Einstein action with a primordial
cosmological constant $\Lambda$, the $B$-terms correspond to the
contribution of the conformal anomaly and the contribution of the
vacuum (Casimir) energy $(B/2a)$ of conformal fields on a static
Einstein spacetime. $F(\eta)$ is the free energy of these fields ---
a typical boson or fermion sum over field oscillators with energies
$\omega$ on a unit 3-sphere, $\eta$ playing the role of the inverse
temperature --- an overall circumference of the toroidal instanton
measured in units of the conformal time. The constant $B$,
    \begin{eqnarray}
    B=\frac{3\beta}{4 m_P^2}=\frac{\beta G}\pi,   \label{B}
    \end{eqnarray}
is determined by the coefficient $\beta$ of the topological
Gauss-Bonnet invariant $E = R_{\mu\nu\alpha\gamma}^2-4R_{\mu\nu}^2 +
R^2$ in the overall conformal anomaly of quantum fields
    \begin{equation}
    g_{\mu\nu}\frac{\delta
    \varGamma_E}{\delta g_{\mu\nu}} =
    \frac{1}{4(4\pi)^2}g^{1/2}
    \left(\alpha \Box R +
    \beta E + \gamma C_{\mu\nu\alpha\beta}^2\right)    \label{anomaly}
    \end{equation}
($C^2_{\mu\nu\alpha\beta}$ is the Weyl tensor squared term). For a
model with $N_0$ scalars, $N_{1/2}$ Weyl spinors and $N_{1}$ gauge
vector fields it reads \cite{Duffanomaly}
    \begin{eqnarray}
    \beta=\frac1{360}\,\big(2 N_0+11 N_{1/2}+
    124 N_{1}\big).                \label{100}
    \end{eqnarray}

The coefficient $\gamma$ does not contribute to (\ref{FrieEu})
because the Weyl tensor vanishes for any FRW metric. What concerns
the coefficient $\alpha$ is more complicated. A nonvanishing
$\alpha$ induces higher derivative terms $\sim \alpha (a'')^2$ in
the action and, therefore, adds one extra degree of freedom to the
minisuperspace sector of $a$ and $N$ and results in
instabilities\footnote{In Einstein theory this sector does not
contain physical degrees of freedom at all, which solves the problem
of the formal ghost nature of $a$ in the Einstein Lagrangian.
Addition of higher derivative term for $a$ does not formally lead to
a ghost -- the additional degree of freedom has a good sign of the
kinetic term as it happens in $f(R)$-theories, but still leads to
the instabilities discovered in \cite{Starobinsky}.}. But $\alpha$
can be renormalized to zero by adding a finite {\em local}
counterterm $\sim R^2$ admissible by the renormalization theory. We
assume this {\em number of degrees of freedom preserving}
renormalization to keep theory consistent both at the classical and
quantum levels \cite{slih}. It is interesting that this finite
renormalization changes the value of the Casimir energy of conformal
fields in closed Einstein cosmology in such a way that for all spins
this energy is universally expressed in terms of the same conformal
anomaly coefficient $B$ (corresponding to the $B/2a$ term in
(\ref{FrieEu})) \cite{slih}. As we will see, this leads to the
gravitational screening of the Casimir energy, mediated by the
conformal anomaly of quantum fields.

Ultimately, the effective action (\ref{FrieEu}) contains only two
dimensional constants -- the Planck mass squared (or the
gravitational constant) $m_P^2=3\pi/4G$ and the cosmological
constant $\Lambda$. They have to be considered as renormalized
quantities. Indeed, the effective action of conformal fields
contains divergences, the quartic and quadratic ones being absorbed
by the renormalization of the initially singular bare cosmological
and gravitational constants to yield finite renormalized $m_P^2$ and
$\Lambda$ \cite{Buchbinder}. Logarithmically divergent counterterms
have the same structure as curvature invariants in the anomaly
(\ref{anomaly}). When integrated over the spacetime closed toroidal
FRW instantons they identically vanish because the $\Box R$ term is
a total derivative, the Euler number $E$ of $S^3\times S^1$ is zero,
$\int d^4x g^{1/2}E=0$, and $C_{\mu\nu\alpha\beta}=0$. There is
however a finite tail of these vanishing logarithmic divergences in
the form of the conformal anomaly action which incorporates the
coefficient $\beta$ of $E$ in (\ref{anomaly}) and constitutes a
major contribution to $\varGamma_E$ --- the first two $B$-dependent
terms of (\ref{effaction})\footnote{These terms can be derived from
the metric-dependent Riegert action \cite{Riegert} or the action in
terms of the conformal factor relating two metrics
\cite{FrTs,BMZ,Buchbinder2} and generalize the action of \cite{FHH}
to the case of a spatially closed cosmology with $\alpha=0$.}. Thus,
in fact, this model when considered in the leading order of the
$1/N$-expansion (therefore disregarding loop effects of the graviton
and other non-conformal fields) is renormalizable in the
minisuperspace sector of the theory.

\begin{figure}[h]
\centerline{\epsfxsize 6cm \epsfbox{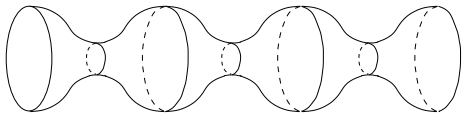}} \caption{\small The
garland segment consisting of three folds of a simple instanton
glued at surfaces of a maximal scale factor.
 \label{Fig.3}}
\end{figure}

The path integral (\ref{1}) is dominated by the saddle points ---
solutions of the equation $\delta\varGamma_E/\delta N(\tau)=0$ which
reads as
    \begin{eqnarray}
    &&-\frac{a'^2}{a^2}+\frac{1}{a^2}
    -B \left(\frac12\,\frac{a'^4}{a^4}
    -\frac{a'^2}{a^4}\right) =
    \frac\Lambda3+\frac{C}{ a^4},     \label{efeq}
    \end{eqnarray}
    with $C$ given by
    \begin{eqnarray}
    &&C = \frac{B}2 +\frac{dF(\eta)}{d\eta},\,\,\,\,
    \eta = 2k \int_{\tau_-}^{\tau_+}
    \frac{d\tau}{a}.                       \label{bootstrap}
    \end{eqnarray}
Note that the usual (Euclidean) Friedmann equation is modified by
the anomalous $B$-term and the radiation term $C/a^4$. The constant
$C$ sets the amount of radiation and satisfies the bootstrap
equation (\ref{bootstrap}), where $B/2$ is the  contribution of the
Casimir energy, and
    \begin{eqnarray}
    \frac{dF(\eta)}{d\eta}=
    \sum_\omega\frac{\omega}{e^{\omega\eta}\mp 1}        \label{100a}
    \end{eqnarray}
is the energy of the gas of thermally excited particles with the
inverse temperature $\eta$. The latter is given in (\ref{bootstrap})
by the $k$-fold integral between the turning points of the scale
factor history $a(\tau)$, $\dot a(\tau_\pm)=0$. This $k$-fold nature
implies that in the periodic solution the scale factor oscillates
$k$ times between its maximum and minimum values
$a_\pm=a(\tau_\pm)$, see Fig.\ref{Fig.3}

As shown in \cite{slih}, such solutions represent garland-type
instantons which exist only in the limited range of values of the
cosmological constant
    \begin{eqnarray}
    0<\Lambda_{\rm min}<\Lambda<
    \frac{3\pi}{2\beta G},                \label{landscape}
    \end{eqnarray}
and eliminate the vacuum Hartle-Hawking instantons corresponding to
$a_-=0$. Hartle-Hawking instantons are ruled out in the statistical
sum by their infinite positive effective action which is due to the
contribution of the conformal anomaly. Hence the tree-level
predictions of the theory are drastically changed.

The upper bound of the range (\ref{landscape}) is entirely caused by
the quantum anomaly -- it represents a new quantum gravity scale
which tends to infinity when one switches the quantum effects off,
$\beta\to 0$. The lower bound $\Lambda_{\rm min}$  can be attributed
to both radiation and anomaly, and can be obtained numerically for
any field content of the model. For a large number of conformal
fields, and therefore a large $\beta$, the lower bound is of the
order $\Lambda_{\rm min}\sim 1/\beta G$. Thus the restriction
(\ref{landscape}) can be regarded as a solution of the cosmological
constant problem in the early Universe, because specifying a
sufficiently high number of conformal fields one can achieve a
primordial value of $\Lambda$ well below the Planck scale where the
effective theory applies, but high enough to generate a sufficiently
long inflationary stage. Also this restriction can be potentially
considered as a selection criterion for the landscape of string
vacua \cite{slih,why}.

\section{Lorentzian quantum gravity density matrix: sum
over Everything}

The period of the quasi-thermal instantons is not a freely
specifiable parameter and can be obtained as a function of $G$ and
$\Lambda$ from Eqs. (\ref{efeq})-(\ref{bootstrap}). Therefore this
model clearly does not describe a canonical ensemble, but rather a
microcanonical ensemble \cite{why} with only two freely specifiable
dimensional parameters  --- the renormalized gravitational and
renormalized cosmological constants as discussed above.

To show this, contrary to the EQG construction of the above type,
consider the density matrix as the canonical path integral in {\em
Lorentzian} quantum gravity. Its kernel in the representation of
3-metrics and matter fields denoted below as $q$ reads
    \begin{eqnarray}
    \rho(q_+,q_-)=e^{\varGamma}\!\!\!\!\!\!\!\!
    \int\limits_{\,\,\,\,\,
    q(t_\pm)=\,q_\pm}
    \!\!\!\!\!\!\!
    D[\,q,p,N\,]\;
    e^{\,i\!\int_{\,\,t_-}^{t_+} dt\,
    (p\,\dot q-N^\mu H_\mu)},             \label{rho}
    \end{eqnarray}
where the integration runs over histories of phase-space variables
$(q(t),p(t))$ interpolating between $q_\pm$ at $t_\pm$ and the
Lagrange multipliers of the gravitational constraints
$H_\mu=H_\mu(q,p)$ --- lapse and shift functions $N(t)=N^\mu(t)$.
The measure $D[\,q,p,N\,]$ includes the gauge-fixing factor of the
delta function $\delta\,[\,\chi\,]=\prod_t\prod_\mu\delta(\chi^\mu)$
of gauge conditions $\chi^\mu$ and the relevant ghost factor
\cite{can,BarvU} (condensed
index $\mu$ includes also continuous spatial labels). 
It is important that the integration range of $N^\mu$,
    \begin{eqnarray}
    -\infty<N<+\infty,    \label{Nrange}
    \end{eqnarray}
generates in the integrand the delta-functions of the constraints
$\delta(H)=\prod_\mu \delta(H_\mu)$. As a consequence the kernel
(\ref{rho}) satisfies the set of Wheeler-DeWitt equations
    \begin{eqnarray}
    \hat H_\mu\big(q,\partial/i\partial q\big)\,
    \rho(\,q,q')=0,             \label{WDW}
    \end{eqnarray}
and the density matrix (\ref{rho}) can be regarded as an operator
delta-function of these constraints
    \begin{eqnarray}
    \hat\rho\sim
    ``\prod_\mu \delta(\hat H_\mu)"
    .        \label{micro}
    \end{eqnarray}
This expression should not be understood literally because the
multiple delta-function here is not uniquely defined, for the
operators $\hat H_\mu$ do not commute
and form an open algebra. Moreover, exact operator realization $\hat
H_\mu$ is not known except the first two orders of a semiclassical
$\hbar$-expansion \cite{geom}. Fortunately, we do not need a precise
form of these constraints, because we will proceed with their
path-integral solutions adjusted to the semiclassical perturbation
theory.

The very essence of our proposal is the interpretation of
(\ref{rho}) and (\ref{micro}) as the density matrix of a {\em
microcanonical} ensemble in spatially closed quantum cosmology. A
simplest analogy is the density matrix of an unconstrained system
having a conserved Hamiltonian $\hat H$ in the microcanonical state
with a fixed energy $E$, $\hat\rho\sim \delta(\hat H-E)$. A major
distinction of (\ref{micro}) from this case is that spatially closed
cosmology does not have freely specifiable constants of motion like
the energy or other global charges. Rather it has as constants of
motion the Hamiltonian and momentum constraints $H_\mu$, all having
a particular value --- zero. Therefore, the expression (\ref{micro})
can be considered as a most general and natural candidate for the
quantum state of the {\em closed} Universe. Below we confirm this
fact by showing that in the physical sector the corresponding
statistical sum is a uniformly distributed (with a unit weight)
integral over entire phase space of true physical degrees of
freedom. Thus, this is the sum over Everything. However, in terms of
the observable quantities, like spacetime geometry, this
distribution turns out to be nontrivially peaked around a particular
set of universes. Semiclassically this distribution is given by the
EQG density matrix and the saddle-point instantons of the above type
\cite{slih}.

From the normalization of the density matrix in the physical Hilbert
space we have
    \begin{eqnarray}
    &&1=
    {\rm Tr_{phys}}\,\hat\rho=\int dq\,
    \mu\big(q,\partial/i\partial
    q \big)\,\rho(q,q')\Big|_{\,q'=q}
    =e^{\,\varGamma}\!\!\!
    \int\limits_{\,\,\rm periodic}
    \!\!\!\!D[\,q,p,N\,]\;
    e^{\,i\int dt (p\,\dot q-N^\mu H_\mu)}.  \label{5}
    \end{eqnarray}
Here in view of the coincidence limit $q'=q$ the integration runs
over periodic histories $q(t)$, and $\mu\big(q,\partial/i\partial
q\big)=\hat\mu$ is the measure which distinguishes the physical
inner product in the space of solutions of the Wheeler-DeWitt
equations $(\psi_1|\psi_2)=\langle\psi_1|\hat\mu|\psi_2\rangle$ from
that of the space of square-integrable functions,
$\langle\psi_1|\psi_2\rangle=\int dq\,\psi_1^*\psi_2$. This measure
includes the delta-function of unitary gauge conditions
$\chi^\mu=\chi^\mu(q,p)$ and an operator factor incorporating the
relevant ghost determinant
\cite{geom}.

On the other hand, in terms of the physical phase space variables
the Faddeev-Popov path integral equals \cite{can,BarvU}
    \begin{eqnarray}
    &&\!\!\!
    \int\limits_{\,\,\rm periodic}
    \!\!\!\!
    D[\,q,p,N\,]\;
    e^{\,i\!\int dt\,(p\,\dot q-N^\mu
    H_\mu)}=\!\!\!
    \int\limits_{\,\,\rm periodic}
    \!\!\!\!
    Dq_{\rm phys}\,Dp_{\rm phys}\,
    e^{i\int dt\,\left(p_{\rm phys}\,
    \dot q_{\rm phys}-H_{\rm phys}(t)\right)}\nonumber\\
    &&\qquad\qquad\qquad\qquad=
    {\rm Tr_{phys}}\,\left(\mathbf{T}\,e^{-i\int dt\,
    \hat H_{\rm phys}(t)}\right) ,             \label{6}
    \end{eqnarray}
where $\mathbf{T}$ denotes the chronological ordering. The physical
Hamiltonian and its operator realization $\hat H_{\rm phys}(t)$ are
nonvanishing here only in unitary gauges explicitly depending on
time \cite{geom}, $\chi^\mu(q,p,t)$. In static gauges,
$\partial_t\chi^\mu=0$, they vanish, because the full Hamiltonian in
closed cosmology is a combination of constraints.

The path integral (\ref{6}) is gauge-independent on-shell
\cite{can,BarvU} and coincides with that in the static gauge.
Therefore, from Eqs.(\ref{5})-(\ref{6}) with $\hat H_{\rm phys}=0$,
the statistical sum of our microcanonical ensemble equals
    \begin{eqnarray}
    &&e^{-\varGamma}={\rm Tr_{phys}}\,\mathbf{I}_{\rm phys}
    =\int dq_{\rm phys}\,dp_{\rm phys}
    ={\rm sum\,\,over\,\,Everything}.     \label{everything}
    \end{eqnarray}
Here $\mathbf{I}_{\rm phys}=\delta(q_{\rm phys}-q_{\rm phys}')$ is a
unit operator in the physical Hilbert space, whose kernel when
represented as a Fourier integral yields extra momentum integration
($2\pi$-factor included into $dp_{\rm phys}$). This sum over
Everything (as a counterpart to the concept of creation from
``anything" in \cite{Star}), not weighted by any nontrivial density
of states, is a result of general covariance and closed nature of
the Universe lacking any freely specifiable constants of motion. The
Liouville integral over entire {\em physical} phase space, whose
structure and topology is not known, is very nontrivial. However,
below we show that semiclassically it is determined by EQG methods
and supported by instantons of \cite{slih} spanning a bounded range
of the cosmological constant.

Integration over momenta in (\ref{5}) yields a Lagrangian path
integral with a relevant measure and action
    \begin{eqnarray}
    e^{-\varGamma}=
    \int D[\,q,N\,]\;e^{iS_L[\,q,\,N\,]}.             \label{7}
    \end{eqnarray}
As in (\ref{5}) integration runs over periodic fields (not indicated
explicitly but assumed everywhere below) even for the case of an
underlying spacetime with Lorentzian signature. Similarly to the
decomposition (\ref{decomposition}) of \cite{slih,lcb} leading to
(\ref{1})-(\ref{2}), we decompose $[\,q,N\,]$ into a minisuperspace
$[\,a_L(t),N_L(t)\,]$ and the ``matter" $\varPhi_L(x)$ variables,
the subscript $L$ indicating their Lorentzian nature. With a
relevant decomposition of the measure
$D[\,q,N\,]=D[\,a_L,N_L\,]\times D\varPhi_L(x)$, the microcanonical
sum reads
    \begin{eqnarray}
    &&\!\!\!\!e^{-\varGamma}=\int\limits
    D[\,a_L,N_L\,]\;
    e^{i\varGamma_L[\,a_L,\,N_L\,]},            \label{3}\\
    &&\!\!\!\!e^{i\varGamma_L[\,a_L,\,N_L\,]}=
    \int\limits
    D\varPhi_L(x)\;
    e^{iS_L[\,a_L,\,N_L;\,\varPhi_L(x)]},   \label{4}
    \end{eqnarray}
where $\varGamma_L[\,a_L,\,N_L\,]$ is a Lorentzian effective action.
The stationary point of (\ref{3}) is a solution of the effective
equation $\delta\varGamma_L/\delta N_L(t)=0$. In the gauge $N_L=1$
it reads as a Lorentzian version of the Euclidean equation
(\ref{efeq}) and the bootstrap equation for the radiation constant
$C$ with the Wick rotated $\tau=it$, $a(\tau)=a_L(t)$, $\eta=i\int
dt/a_L(t)$. However, with these identifications $C$ turns out to be
purely imaginary (in view of the complex nature of the free energy
$F(i\!\int dt/a_L)$). Therefore, no periodic solutions exist in
spacetime with a {\em real} Lorentzian metric.

On the contrary, such solutions exist in the Euclidean spacetime.
Alternatively, the latter can be obtained with the time variable
unchanged $t=\tau$, $a_L(t)=a(\tau)$, but with the Wick rotated
lapse function
    \begin{eqnarray}
    N_L=-i N,\,\,\,\,
    iS_L[\,a_L,N_L;\phi_L]=
    -S_E[\,a,N;\varPhi\,].       \label{8}
    \end{eqnarray}
In the gauge $N=1$ $(N_L=-i)$ these solutions exactly coincide with
the instantons of \cite{slih}. The corresponding saddle points of
(\ref{3}) can be attained by deforming the integration contour
(\ref{Nrange}) of $N_L$ into the complex plane to pass through the
point $N_L=-i$ and relabeling the real Lorentzian $t$ with the
Euclidean $\tau$. In terms of the Euclidean $N(\tau)$, $a(\tau)$ and
$\varPhi(x)$ the integrals (\ref{3}) and (\ref{4}) take the form of
the path integrals (\ref{1}) and (\ref{2}) in EQG,
    \begin{eqnarray}
    i\varGamma_L[\,a_L,\,N_L]=-\varGamma_E[\,a,\,N\,]. \label{9}
    \end{eqnarray}
However, the integration contour for the Euclidean $N(\tau)$ runs
from $-i\infty$ to $+i\infty$ through the saddle point $N=1$. This
is the source of the conformal rotation in Euclidean quantum
gravity, which is called to resolve the problem of unboundedness of
the gravitational action and effectively renders the instantons a
thermal nature, even though they originate from the microcanonical
ensemble. This mechanism implements the justification of EQG from
the canonical quantization of gravity \cite{HSch} (see also
\cite{BYork} for the black hole context).

\section{Cosmological evolution from the initial microcanonical state:
origin of inflation and standard GR scenario}

The gravitational instantons of Sect.2 can be regarded as setting
initial conditions for the cosmological evolution in the physical
spacetime with the Lorentzian signature. Indeed,  those initial
conditions can be viewed as those at the nucleation of the
Lorentzian spacetime from the Euclidean spacetime at the maximum
value of the scale factor $a_+=a(\tau_+)$ at the turning point of
the Euclidean solution $\tau_+$ --- the minimal (zero extrinsic
curvature) surface of the instanton. For the contribution of the
one-fold instanton to the density matrix of the Universe this
nucleation process is depicted in Fig. \ref{Fig.1}.

The Lorentzian evolution can be obtained by analytically continuing
the Euclidean time into the complex plane by the rule
$\tau=\tau_++it$. Correspondingly the Lorentzian effective equation
follows from the Euclidean one (\ref{efeq}) as
    \begin{eqnarray}
    \frac{\dot a^2}{a^2}+\frac{1}{a^2}-
    \frac{B}2 \left(\frac{\dot a^2}{a^2}
    +\frac{1}{a^2}\right)^2 =
    \frac\Lambda3+\frac{C-B/2}{ a^4},   \label{Friedmann3}
    \end{eqnarray}
where the dot, from now on, denotes the derivative with respect to
the Lorentzian time $t$. This can be solved for the Hubble factor as
    \begin{eqnarray}
    &&\frac{\dot
    a^2}{a^2}+\frac{1}{a^2}=
    \frac1B\left\{1-
    \sqrt{1-2B\left(\frac\Lambda3
    +\frac{\cal C}{a^4}\right)}\right\},         \label{Friedmann0}\\
    &&{\cal C} \equiv C-\frac{B}2.              \label{calC}
    \end{eqnarray}
We have thus obtained a modified Friedmann equation in which the
overall energy density, including both the cosmological constant and
radiation, very nonlinearly contributes to the square of the Hubble
factor \cite{BDK}.

An interesting property of this equation is that the Casimir energy
does not weigh. Indeed the term $B/2a^4$ is completely subtracted
from the full radiation density $C/a^4$ in the right hand side of
(\ref{Friedmann3}) and under the square root of (\ref{Friedmann0}).
Only ``real" thermally excited quanta contribute to the right-hand
side of (\ref{Friedmann0}). Indeed, using (\ref{bootstrap}), the
radiation contribution ${\cal C}/a^4$ is seen to read simply as
    \begin{eqnarray}
    \frac{\cal C}{a^4} = \frac1{a^4}
    \sum_\omega\frac{\omega}{e^{\omega\eta}\mp 1}. \label{primrad}
    \end{eqnarray}
This is an example of the gravitational screening which is now being
intensively searched for the cosmological constant
\cite{WoodardTsamis,DvaliKhouryetal}. As we see, in our case, this
mechanism is mediated by the conformal anomaly action, but it
applies not to the cosmological constant, but rather to the Casimir
energy which has the equation of state of radiation
$p=\varepsilon/3$. This gravitational screening is essentially based
on the above mentioned renormalization that eradicates higher
derivatives from the effective action and thus preserves the
minisuperspace sector free from dynamical degrees of freedom.

After nucleation from the Euclidean instanton at the turning point
with $a=a_+$ and $\dot a_+=0$ the Lorentzian Universe starts
expanding, because $\ddot a_+>0$. Therefore, the radiation quickly
dilutes, so that the primordial cosmological constant starts
dominating and can generate an inflationary stage. It is natural to
assume that the primordial $\Lambda$ is not fundamental, but is due
to some inflaton field. This effective $\Lambda$ is nearly constant
during the Euclidean stage and the inflation stage, and subsequently
leads to a conventional exit from inflation by the slow roll
mechanism\footnote{In the Euclidean regime this field also stays in
the slow roll approximation, but in view of the oscillating nature
of a scale factor it does not monotonically decay. Rather it follows
these oscillations with much lower amplitude and remains nearly
constant during all Euclidean evolution, whatever long this
evolution is (as it happens for garland instantons with the number
of folds $k\to\infty$).}.

During a sufficiently long inflationary stage, particle production
of conformally non-invariant matter takes over the polarization
effects of conformal fields. After being thermalized at the exit
from inflation this matter gives rise to an energy density
$\varepsilon(a)$ which should replace the energy density of the
primordial cosmological constant and radiation. Therefore, at the
end of inflation the combination $\Lambda/3+{\cal C}/a^4$ should be
replaced according to
    \begin{eqnarray}
    \frac\Lambda3+\frac{\cal C}{a^4}\to
    \frac{8\pi G}3\,\varepsilon(a)\equiv
    \frac{8\pi G}3\,\rho(a)+\frac{\cal C}{a^4}.   \label{split}
    \end{eqnarray}
Here $\varepsilon(a)$ denotes the full energy density including the
component $\rho(a)$ resulting from the decay of $\Lambda$ and the
radiation density of the primordial conformal matter ${\cal C}/a^4$.
The dependence of $\varepsilon(a)$ on $a$ is of course determined by
the equation of state via the stress tensor conservation, and
$\rho(a)$ also includes its own radiation component emitted by and
staying in (quasi)equilibrium with the
 baryonic part of the full $\varepsilon(a)$.

Thus the modified Friedmann equation finally takes the form
\cite{BDK}
    \begin{eqnarray}
    \frac{\dot a^2}{a^2}+\frac{1}{a^2}=
    \frac\pi{\beta G}\left\{\,1-
    \sqrt{\,1-\frac{16 G^2}3\,
    \beta\varepsilon}\,\right\},              \label{modFriedmann}
    \end{eqnarray}
where we expressed $B$ according to (\ref{B}).

In the limit of small subplanckian energy density $\beta
G^2\varepsilon\equiv\beta\varepsilon/\varepsilon_P\ll 1$ the
modified equation goes over into the ordinary Friedmann equation in
which the parameter $\beta$ completely drops out
    \begin{eqnarray}
    \frac{\dot a^2}{a^2}+\frac{1}{a^2}
    =\frac{8\pi
    G}3\,\varepsilon.     \label{GR}
    \end{eqnarray}
Therefore within this energy range the standard cosmology is
recovered. Depending on the effective equation of state, a wide set
of the  standard scenarios of late cosmological evolution can be
obtained, including those showing a cosmic acceleration, provided
some kind of dark energy component is present \cite{dark,otherDE}.

\section{Cosmological acceleration -- Big Boost scenario}

The range of applicability of the GR limit (\ref{GR}) depends on
$\beta$. This makes possible a very interesting mechanism to happen
for a very large $\beta$. Indeed, the value of the argument of the
square root in (\ref{modFriedmann}) can be sufficiently far from 1
even for small $\varepsilon$ provided $\beta\sim N_{\rm cdf}\gg 1$.
Moreover, one can imagine a model with a variable number of
conformal fields $N_{\rm cdf}(t)$ inducing a time-dependent, and
implicitly a scale factor-dependent $\beta$, $\beta=\beta(a)$. If
$\beta(a)$ grows with $a$ faster than the rate of decrease of
$\varepsilon(a)$, then the solution of (\ref{modFriedmann}) can
reach a singular point, labeled below by $\infty$, at which the
square root argument vanishes and the cosmological acceleration
becomes infinite. This follows from the expression
    \begin{eqnarray}
    \frac{\ddot a}a\sim\frac{4\pi}{3\beta G}
    \frac{a(G^2\beta\varepsilon)'}{\sqrt{\,1-16 G^2
    \beta\varepsilon/3}},     \label{acceleration}
    \end{eqnarray}
where prime denotes the derivative with respect to $a$. This
expression becomes singular at $t=t_\infty$ even though the Hubble
factor $H^2\equiv ({\dot a^2}/{a^2}+1/{a^2})$ remains finite when
    \begin{eqnarray}
    (G^2\beta\varepsilon)_\infty=\frac3{16},\,\,\,
    H^2_\infty =\frac\pi{\beta G}.              \label{Hinfinity}
    \end{eqnarray}

Assuming for simplicity that the matter density has a dust-like
behavior and $\beta$ grows as a power law in $a$
    \begin{eqnarray}
    G\varepsilon\sim \frac1{a^3},\,\,\,\,
    G\beta\sim
    a^n,\,\,\,\,n>3,              \label{behavior}
    \end{eqnarray}
one easily finds an inflection point $t=t_*$ when the cosmological
acceleration starts after the deceleration stage when
    \begin{eqnarray}
    (G^2\beta\varepsilon)_*=\frac34\frac{n-2}{(n-1)^2},\,\,\,
    H^2_*=\frac{2\pi}{n-1}\frac1{\beta G}.           \label{H*}
    \end{eqnarray}

The evolution ends in this model with the curvature singularity,
$\ddot a\to\infty$, reachable in a finite proper time. Unlike the
situation with a big brake singularity of \cite{bigbrake} it cannot
be extended beyond this singularity analytically even by smearing it
out or taking into account its weak integrable nature. In contrast
to \cite{bigbrake} the acceleration at the singularity is positive.
Hence, we called this type of singularity a {\em big boost}
\cite{BDK}. The effect of the conformal anomaly drives the expansion
of the Universe to the maximum value of the Hubble constant, after
which the solution becomes complex. This, of course, does not make
the model a priori inconsistent, because for $t\to t_\infty$ an
infinitely growing curvature invalidates the semiclassical and $1/N$
approximations. This is a new essentially quantum stage which
requires a UV completion of the effective low-energy theory.

\section{Hierarchy problem, strings and extra dimensions}

As follows from (\ref{landscape}) and (\ref{Hinfinity})-(\ref{H*})
the inflation (which is run by the early primordial $\Lambda$) and
cosmological acceleration stage both have Hubble factors given by
the conformal anomaly coefficient $\beta$, $H^2\sim 1/\beta G$.
Therefore we have to reconcile the inflation and present Dark Energy
scales in the hierarchy
    \begin{eqnarray}
    &&H^2_{\rm inflation}
    =\#\frac{m_P^2}{\beta_{\rm inflation}}\sim ({\rm GUT\; scale})^2,\\
    &&H^2_{\rm present}
    =\#\frac{m_P^2}{\beta_{\rm present}}\sim (10^{-33}{\rm eV})^2,
    \end{eqnarray}
which can be done only by the assumption of the parameter $\beta$
being a time dependent and tremendously growing variable,
$\beta_{\rm present}\gg\beta_{\rm inflation}$. Of course, this can
be considered as a rationale for the Big Boost mechanism of the
previous section. This is not unusual now to suggest a solution to
the hierarchy problem by introducing a gigantic number of quantum
fields, like it has been done in \cite{Dvalispecies} in the form of
$10^{32}$ replicas of the Standard Model resolving the leap between
the electroweak and Planck scales. Whatever speculative are such
suggestions, we would consider a similar mechanism in our case even
though we would have to consider a much bigger jump of about hundred
twenty orders of magnitude.

So what can be the mechanism of a variable and indefinitely growing
$\beta$? One such mechanism was suggested in \cite{why}. It relies
on the possible existence, motivated by string theory, of extra
dimensions whose size is evolving in time. Theories with extra
dimensions can promote $\beta$ to the level of a  modulus variable
which can grow with the evolving size $L$ of those dimensions, as we
now explain. Indeed, the parameter $\beta$ basically counts the
number $N_{\rm cdf}$ of conformal degrees of freedom, $\beta\sim
N_{\rm cdf}$ (see Eq.(\ref{100})). However, if one considers a
string theory in a spacetime with more than four dimensions, the
extra-dimension being compact with typical size $L$, the effective
4-dimensional fields arise as Kaluza-Klein (KK) and winding modes
with masses (see e.g. \cite{Polch})
    \begin{eqnarray}
    m_{n,w}^2=\frac{n^2}{L^2}+\frac{w^2}{\alpha'^2}\,L^2
    \end{eqnarray}
(where $n$ and $w$ are respectively the KK and winding numbers),
which break their conformal symmetry. These modes remain
approximately conformally invariant as long as their masses are much
smaller than the spacetime curvature, $m_{n,w}^2\ll H^2\sim
m_P^2/N_{\rm cdf}$. Therefore the number of conformally invariant
modes changes with $L$. Simple estimates show that the number of
pure KK modes ($w=0$, $n\leq N_{\rm cdf}$) grows with $L$ as $N_{\rm
cdf}\sim (m_P L)^{2/3}$, whereas the number of pure winding modes
($n=0$, $w\leq N_{\rm cdf}$) grows as $L$ decreases as $N_{\rm
cdf}\sim(m_P\alpha'/L)^{2/3}$. Thus, it is possible to find a
growing $\beta$ in both cases with expanding or contracting extra
dimensions. In the first case it is the growing tower of
superhorizon KK modes which {\it makes} the horizon scale
    \begin{eqnarray}
    H\sim \frac{m_P}{\sqrt{N_{\rm cdf}}}\sim\frac{m_P}{(m_P
    L)^{1/3}}
    \end{eqnarray}
decrease as $L$ increases to infinity. In the second case it is the
tower of superhorizon winding modes which makes this scale decrease
with the decreasing $L$ as
    \begin{eqnarray}
    H\sim m_P\left(\frac{L}{m_P\alpha'}\right)^{1/3}.
    \end{eqnarray}
At the qualitative level of this discussion, such a scenario is
flexible enough to accommodate the present day acceleration scale
(though, at the price of fine-tuning an enormous coefficient
governing the expansion or contraction of $L$).

\section{Dual description via the AdS/CFT correspondence}

String (or rather string-inspired) models can offer a more explicit
construction of these ideas within the AdS/CFT picture. Indeed, in
this picture \cite{AdS/CFT} a higher dimensional theory of gravity,
namely type IIB supergravity compactified on $AdS_5 \times S^5$, is
seen to be equivalent to a four dimensional conformal theory, namely
${\cal N}=4$ $SU(N)$ SYM, thought to live on the boundary of $AdS_5$
space-time. This picture underlies the Randall-Sundrum model
\cite{Randall:1999vf} where a 3-brane embedded into the $AdS_5$
space-time enjoys in a large distance limit a recovery of the 4D
gravity theory without the need for compactification \cite{Gubser}.
This model has a dual description. On the one hand it can be
considered from a 5D gravity perspective, on the other hand it can
also be described, thanks to the AdS/CFT correspondence, by a 4D
conformal field theory coupled to gravity.

To be more precise, the 5D SUGRA --- a field-theoretic limit of
compactified type IIB string theory --- induces on the  brane  of
the underlying AdS background the quantum effective action of the
conformally invariant 4D ${\cal N}=4$ $SU(N)$ SYM theory coupled to
the 4D geometry of the boundary. The multiplets of this CFT
contributing according to (\ref{100}) to the total conformal anomaly
coefficient $\beta$ are given by $(N_0,N_{1/2},N_1)=(6N^2,4N^2,N^2)$
\cite{DuffLiu}, so that
    \begin{eqnarray}
    \beta=\frac12\,N^2.
    \end{eqnarray}
The parameters of the two theories are related by the equation
\cite{AdS/CFT,Gubser,HHR}
    \begin{eqnarray}
    \frac{L^3}{2 G_5}=\frac{N^2}{\pi},
    \end{eqnarray}
where $L$ is the radius of the 5D $AdS$ space-time with the negative
cosmological constant $\Lambda_5=-6/L^2$ and $G_5$ is the 5D
gravitational constant. The radius $L$ is also related to the 't
Hooft parameter of the SYM coupling $\lambda=g_{SYM}^2 N$ and the
string length scale $l_s=\sqrt{\alpha'}$, $L=\lambda^{1/4} l_s$. The
generation of the 4D CFT from the local 5D supergravity holds in the
limit when both $N$ and $\lambda$ are large. This guarantees the
smallness of string corrections and establishes the relation between
the weakly coupled tree-level gravity theory in the bulk ($G_5\to
0$, $L\to\infty$) and the strongly coupled 4D CFT ($g_{SYM}^2\gg
1$). Moreover, as said above, the AdS/CFT correspondence explains
the mechanism of recovering general relativity theory on the 4D
brane of the Randall-Sundrum model \cite{Gubser,HHR}. The 4D gravity
theory is induced on the brane from the 5D theory with the negative
cosmological constant $\Lambda_5=-6/L^2$. In the one-sided version
of this model the brane has a tension $\sigma=3/8\pi G_5L$ (the 4D
cosmological constant is given by $\Lambda_4=8\pi G_4\sigma$), and
the 4D gravitational constant $G_4\equiv G$ turns out to be
    \begin{eqnarray}
    G=\frac{2G_5}L.
    \end{eqnarray}
One recovers 4D General Relativity at low energies and for distances
larger than the radius of the AdS bulk, $L$. Thus, the CFT dual
description of the 5D Randall-Sundrum model is very similar to the
model considered above. Moreover, even though the CFT effective
action is not exactly calculable for $g_{SYM}^2\gg 1$ it is
generally believed that its conformal anomaly is protected by
extended SUSY \cite{TseytlinLiu} and is exactly given by the
one-loop result (\ref{anomaly}). Therefore it generates the exact
effective action of the anomalous (conformal) degree of freedom
given by (\ref{effaction}), which guarantees a good $1/N_{\rm
cdf}$-approximation for the gravitational dynamics.

Applying further the above relations it follows a relation between
our $\beta$ coefficient and the radius $L$ of the $AdS$ space-time,
given by $\beta G=\pi L^2/2$. Introducing this in the modified
Friedmann equation (\ref{modFriedmann}), the latter becomes
explicitly depending on the size of the 5D AdS spacetime as given by
    \begin{eqnarray}
    \frac{\dot a^2}{a^2}+\frac{1}{a^2}=
    \frac2{L^2}\left\{\,1-
    \sqrt{\,1-L^2 \left(\frac{8\pi G}3\,
    \rho+\frac{\cal C}{a^4}\right)}\,\right\}, \label{modFriedmann2}
    \end{eqnarray}
where we have reintroduced the decomposition (\ref{split}) of the
full matter density into the decay product of the inflationary and
matter domination stages, with energy density $\rho$, and the
thermal excitations of the primordial CFT (\ref{primrad}).

For low energy density, $GL^2\rho\ll 1$ and $L^2 {\cal C}/a^4\ll 1$,
in the approximation beyond the leading order, cf. Eq.(\ref{GR}),
the modified Friedmann equation coincides with the modified
Friedmann equation in the Randall-Sundrum model \cite{BinDefLan}
    \begin{eqnarray}
    \frac{\dot a^2}{a^2}+\frac{1}{a^2}
    =\frac{8\pi
    G}3\,\rho\,
    \left(1+\frac{\rho}{2\sigma}\right)+
    \frac{\cal C}{a^4},                      \label{RS}
    \end{eqnarray}
where $\sigma=3/8\pi G_5L=3/4\pi GL^2$ is the Randall-Sundrum brane
tension and ${\cal C}$ is the braneworld constant of motion
\cite{BinDefLan,bulkBH}.\footnote{We assume that the dark radiation
term is redshifted, as $a$ grows, faster than the matter term and
expand to the second order in $\rho$, but the first order in ${\cal
C}$.} Note that the thermal radiation on the brane (of non-Casimir
energy nature) is equivalent to the mass of the bulk black hole
associated with this constant. This fact can be regarded as another
manifestation of the AdS/CFT correspondence in view of the known
duality between the bulk black hole and the thermal CFT on the brane
\cite{bulkBH}.

Interestingly, this comparison between our model and the
Randall-Sundrum framework also allows  one to have some insight on
the phenomenologically allowed physical scales. Indeed, it is well
known that the presence of an extra-dimension in the Randall-Sundrum
model, or in the dual language, that of the CFT, manifests itself
typically at distances lower than the $AdS$ radius $L$. Hence, it is
perfectly possible to have a large number of conformal fields in the
Universe, {\it \`a  la} Randall-Sundrum, without noticing their
presence in the everyday experiments, provided $L$ is small enough.
Moreover, if one uses the scenario of \cite{slih} to set the initial
conditions for inflation, it provides an interesting connection
between the Hubble radius of inflation, given by eq.
(\ref{landscape}), and the distance at which the presence of the CFT
would manifest itself in gravity experiments, both being given by
$L$. Last, it seems natural in a string theory setting, to imagine
that the $AdS$ radius $L$ can depend on time, and hence on the scale
factor.

In this case, assuming that the AdS/CFT picture still holds when $L$
is adiabatically evolving, one can consider the possibility that
$GL^2\varepsilon$ is large, and that $L^2(t)$ grows faster than
$G\varepsilon(t)$ decreases during the cosmological expansion. One
would then get the cosmological acceleration scenario of the above
type followed by the big boost singularity.

In this case, however, should this acceleration scenario correspond
to the present day accelerated expansion, $L$ should be of the order
of the present size of the Universe, i.e. $L^{-2}\sim H_{\rm
present}^2$. Since the Randall-Sundrum mechanism recovers 4D GR only
at distances beyond the curvature radius of the AdS bulk, $r\gg L$,
it means that local gravitational physics of our model
(\ref{modFriedmann2}) at the acceleration stage is very different
from the 4D general relativity. Thus this mechanism can hardly be a
good candidate for generating dark energy in real cosmology.

\section{Anomaly driven cosmology and the DGP model}

It is interesting that there exists an even more striking example of
a braneworld setup dual to our anomaly driven model. This is the
generalized DGP model \cite{DGP} including together with the 4D and
5D Einstein-Hilbert terms also the 5D cosmological constant,
$\Lambda_5$, in the special case of the {\em vacuum} state on the
brane with a vanishing matter density $\rho=0$. In contrast to the
Randall-Sundrum model, for which this duality holds only in the low
energy limit --- small $\rho$ and small ${\cal C}/a^4$, vacuum DGP
cosmology {\em exactly} corresponds to the model of \cite{slih} with
the 4D cosmological constant $\Lambda$ simulated by the 5D
cosmological constant $\Lambda_5$.

Indeed, in this model (provided one neglects the bulk curvature),
gravity interpolates between a 4D behaviour at small distances and a
5D behaviour at large distances, with the crossover scale between
the two regimes being given by $r_c$,
    \begin{eqnarray}
    \frac{G_5}{2G}=r_c,      \label{DGPscale}
    \end{eqnarray}
and in the absence of stress-energy exchange between the brane and
the bulk, the modified Friedmann equation takes the form
\cite{DGPDeffayet}
    \begin{eqnarray}
    \frac{\dot a^2}{a^2}+\frac{1}{a^2}-
    r_c^2 \left(\,\frac{\dot a^2}{a^2}
    +\frac{1}{a^2}-\frac{8\pi G}3\,\rho\right)^2 =
    \frac{\Lambda_5}{6}
    +\frac{{\cal C}}{ a^4}.             \label{FriedmannDGP}
    \end{eqnarray}
Here ${\cal C}$ is the same as above constant of integration of the
bulk Einstein's equation, which corresponds to a nonvanishing Weyl
tensor in the bulk (or a mass for a Schwarzschild geometry in the
bulk) \cite{BinDefLan,bulkBH}. It is remarkable that this equation
with $\rho=0$ exactly coincides with the modified Friedmann equation
of the anomaly driven cosmology (\ref{Friedmann3}) under the
identifications
    \begin{eqnarray}
    &&B\equiv\frac{\beta G}\pi=2 r_c^2, \label{1000}\\
    &&\Lambda=\frac{\Lambda_5}2.
    \end{eqnarray}
These identifications imply that in the DGP limit $G\ll r_c^2$, the
anomaly coefficient $\beta$ is much larger than 1.

This looks very much like the generation of the vacuum DGP model for
any value of the dark radiation ${\cal C}/a^4$ from the anomaly
driven cosmology with a very large $\beta\sim m_P^2 r_c^2\gg 1$.
However, there are several differences. A first important difference
between the conventional DGP model and the anomaly driven DGP is
that the former does not incorporate the self-accelerating branch
\cite{DGPDeffayet,DDG} of the latter. This corresponds to the fact
that only one sign of the square root is admissible in
Eq.(\ref{Friedmann0}) --- a property dictated by the instanton
initial conditions at the nucleation of the Lorentzian spacetime
from the Euclidean one. So, one does not have to worry about
possible instabilities associated with the self-accelerating branch.

 Another important difference concerns the way the matter energy
density manifests itself in the Friedmann equation for the
non-vacuum case.  In our 4D anomaly driven model it enters the right
hand side of the equation as a result of the decay (\ref{split}) of
the effective 4D cosmological constant $\Lambda$, while in the DGP
model it appears inside the parenthesis of the left hand side of
equation (\ref{FriedmannDGP}). Therefore, the DGP Hubble factor
reads as
    \begin{eqnarray}
    \frac{\dot a^2}{a^2}+\frac{1}{a^2}=
    \frac{8\pi G}3\,
    \rho+
    \frac1{2r_c^2}\left\{\,1-
    \sqrt{\,1-4r_c^2
    \left(\frac{\textstyle\Lambda_5}{\textstyle 6}
    +\frac{\textstyle\cal C}{\textstyle a^4}-
    \frac{\textstyle 8\pi G}{\textstyle 3}\,
    \rho\right)}\,\right\}                     \label{modFriedmannDGP}
    \end{eqnarray}
(note the negative sign of $\rho$ under the square root and the
extra first term on the right hand side). In the limit of small
$\rho$, ${\cal C}/a^4$ and $\Lambda_5$, the above equation yields a
very different behavior  from the GR limit of the anomaly driven
model (\ref{GR}),
    \begin{eqnarray}
    \frac{\dot a^2}{a^2}+\frac{1}{a^2}\simeq
    \frac{\textstyle\Lambda_5}{\textstyle 6}
    +\frac{\textstyle\cal C}{\textstyle a^4}
    +r_c^2
    \left(\frac{\textstyle\Lambda_5}{\textstyle 6}
    +\frac{\textstyle\cal C}{\textstyle a^4}-
    \frac{\textstyle 8\pi G}{\textstyle 3}\,
    \rho\right)^2.
    \end{eqnarray}
For vanishing $\Lambda_5$ and ${\cal C}/a^4$ this behavior
corresponds to the 5D dynamical phase \cite{DGPDeffayet,DDG} which
is realized in the DGP model for a very small matter energy density
on the brane $\rho\ll 3/32\pi G r_c^2\sim m_P^2/r_c^2$.

Of course, in this range the DGP braneworld reduces to a vacuum
brane, but one can also imagine that the 5D cosmological constant
decays into matter constituents similar to (\ref{split}) and thus
simulates the effect of $\rho$ in Eq.(\ref{modFriedmann}). This can
perhaps provide us with a closer correspondence between the anomaly
driven cosmology and the non-vacuum DGP case. But here we would
prefer to postpone discussions of such scenarios to future analyses
and, instead, focus on the generalized {\em single-branch} DGP model
to show that it also admits the cosmological acceleration epoch
followed by the big boost singularity.

Indeed, for positive $\Lambda_5$ satisfying a very weak bound
    \begin{eqnarray}
    \Lambda_5>\frac3{2r_c^2} \label{bound}
    \end{eqnarray}
Eq.(\ref{modFriedmannDGP}) has a solution for which,  during the
cosmological expansion with $\rho\to 0$, the argument of the square
root vanishes and the acceleration tends to $\pm\infty$. For the
effective $a$-dependence of $r_c^2$ and $G\rho$ analogous to
(\ref{behavior}), $r_c^2(a)\sim a^n$ and $G\rho(a)\sim 1/a^3$, the
acceleration becomes positive at least for $n\geq 0$,
    \begin{eqnarray}
    \frac{\ddot a}a\simeq
    \frac{\textstyle
    n+ 32\pi G\,r_c^2\rho}
    {\textstyle 4r_c^2\,\sqrt{\,1+4r_c^2
    \Big(\frac{\textstyle 8\pi G}{\textstyle 3}\,
    \rho-\frac{\textstyle\Lambda_5}{\textstyle 6}
    -\frac{\textstyle\cal C}{\textstyle a^4}\Big)}}
    \to+\infty.
    \end{eqnarray}
This is the big boost singularity labeled by $\infty$ and having a
finite Hubble factor $({\dot a^2}/{a^2}+1/{a^2})_\infty
=\Lambda_5/6+1/4r_c^2$.

Thus, the {\em single-branch} DGP cosmology can also lead to a big
boost version of acceleration. For that to happen, one does not
actually need a growing $r_c$  (which can be achieved at the price
of having a time dependent $G_5$ --- itself some kind of a modulus,
in a string inspired picture). The DGP crossover scale $r_c$ can be
constant, $n=0$, and the big boost singularity will still occur
provided the lower bound (\ref{bound}) is satisfied \footnote{More
precisely, one should also take into account here the modification
due the dark  radiation contribution ${\cal C}/a^4$. However, the
latter is
 very small at late stages of
expansion.}. When $\Lambda_5$ violates this bound, the acceleration
stage is eternal with an asymptotic value of the Hubble factor
squared $H^2 = \dot a^2/a^2$ given by
$\big(1-\sqrt{1-2r_c^2\Lambda_5/3}\big)/2r_c^2$.

\section{Conclusions}
To summarize, within a minimum set of assumptions (the equipartition
in the physical phase space (\ref{everything})), we have a
mechanism to generate a limited range of a positive cosmological
constant which is likely to constrain the landscape of string vacua
and get the full evolution of the Universe as a quasi-equilibrium
decay of its initial microcanonical state. Thus, contrary to
anticipations of Sidney Coleman that ``there is nothing rather than
something" regarding the actual value of the cosmological constant
\cite{baby}, one can say that something (rather than nothing) comes
from everything.

We have obtained the modified Friedmann equation for this evolution
in the anomaly dominated cosmology. This equation exhibits a
gravitational screening of the quantum Casimir energy of conformal
fields --- this part of the total energy density does not weigh,
being degravitated due to the contribution of the conformal anomaly.
Also, in the low-density limit this equation does not only show a
recovery of the standard general relativistic behavior, but also
coincides with the dynamics of the Randall-Sundrum cosmology within
the AdS/CFT duality relations. Moreover, for a very large and
rapidly growing value of the Gauss-Bonnet coefficient $\beta$ in the
conformal anomaly this equation features a regime of cosmological
acceleration followed by a big boost singularity. At this
singularity the acceleration factor grows in finite proper time up
to infinity with a finite limiting value of the Hubble factor. A
proper description of the late phase of this evolution, when the
Universe enters again a quantum phase, would require a UV completion
of the low-energy semiclassical theory.

A natural mechanism for a growing $\beta$ can be based on the idea
of an adiabatically evolving scale associated with extra dimensions
\cite{why} and realized within the picture of AdS/CFT duality,
according to which a conformal field theory is induced on the 4D
brane from the 5D non-conformal theory in the bulk. As is well
known, this duality sheds light on the 4D general relativistic limit
in the Randall-Sundrum model \cite{Gubser,HHR}. Here we observed an
extended status of this duality from the cosmological perspective
--- the generalized Randall Sundrum model with the
Schwarzschild-AdS bulk is equivalent to the anomaly driven cosmology
for small energy density. In particular, the radiation  content of
the latter is equivalent to the dark radiation term ${\cal C}/a^4$
pertinent to the Randall-Sundrum braneworld with a bulk black hole
of mass ${\cal C}$.

Another intriguing observation concerns the {\em exact}
correspondence between the anomaly driven cosmology and the vacuum
DGP model generalized to the case of a nonvanishing bulk
cosmological constant $\Lambda_5$. In this case a large $\beta$ is
responsible for the large crossover scale $r_c$, (\ref{DGPscale}).
For positive $\Lambda_5$ satisfying the lower bound (\ref{bound})
this model also features a big boost scenario even for stabilized
$\beta$. Below this bound (but still for positive $\Lambda_5>0$,
because a negative $\Lambda_5$ would imply a time of maximal
expansion from which the Universe would start recollapsing) the
cosmological evolution eventually enters an eternal acceleration
phase. However, the DGP model with  matter on the brane can hardly
be equivalent to the 4D anomaly driven cosmology, unless one has
some mechanism for $\Lambda_5$ to decay and to build up matter
density on the brane.

Unfortunately, our scenario put in the framework of the AdS/CFT
correspondence with adiabatically evolving scale of extra dimension
cannot agree with the observed dark energy, because, for the
required values of the parameters, the local gravitational physics
of this model would become very different from the 4D general
relativity.

In general, the idea of a very large central charge of CFT algebra,
underlying the solution of the hierarchy problem in the dark energy
sector and particle phenomenology, seems hovering in current
literature \cite{bigcentalcharge,Dvalispecies}. Our idea of a big
growing $\beta$ belongs to the same scope, but its realization seems
missing a phenomenologically satisfactory framework. In essence, it
can be considered as an attempt to cross a canyon in two endeavors
--- the leap of 32 decimal orders of magnitude in $N_{\rm cdf}$
of \cite{Dvalispecies}, separating the electroweak and Planckian
scales, versus our 120 orders of magnitude needed to transcend
separation between the Hubble and Planckian scales. Both look
equally speculative from the viewpoint of local phenomenology.

Probably some other modification of this idea can be more
productive. In particular, an alternative mechanism of running
$\beta$ could be based on the winding modes. These modes do not seem
to play essential role in the AdS/CFT picture with a big scale of
extra dimensions $L$, because they are heavy in this limit. On the
contrary, this mechanism should work in the opposite case of
contracting extra dimensions, for which the restrictions from local
gravitational physics do not apply (as long as for $L\to 0$ the
short-distance correction go deeper and deeper into UV domain).

\section*{Acknowledgements}
The work of A.O.B. was supported by the Russian Foundation for Basic
Research under the grant No 08-01-00737 and the grant
LSS-1615.2008.2. The collaboration of A.B. and C.D. was made in part
possible via the ANR grant "MODGRAV". A.K. was partially supported by the
RFBR grant 08-02-00923, the grant LSS-4899.2008.2 and by the Research
Programme "Elementary Particles" of the Russian Academy of Sciences.


\begin{thebibliography}{99}
\bibitem{HH}J.B.Hartle and S.W.Hawking, Phys.Rev. {\bf D28}, 2960
(1983); S.W.Hawking, Nucl. Phys. {\bf B 239}, 257 (1984).

\bibitem{tunnel}A.D. Linde, JETP {\bf 60}, 211 (1984);
A.Vilenkin, Phys. Rev. {\bf D 30}, 509 (1984).

\bibitem{baby}S.R.Cole\-man, Nucl. Phys. {\bf B 310}, 643 (1988).

\bibitem{stochastic}
A.A.Starobinsky, in {\em Field Theory, Quantum Gravity and Strings},
107 (eds. H.De Vega and N.Sanchez, Springer, 1986); A.D.Linde, {\em
Particle physics and inflationary cosmology} (Harwood, Chur,
Switzerland, 1990).

\bibitem{GHP}G.W.Gibbons, S.W.Hawking and M.Perry, Nucl. Phys. {\bf
B 138}, 141 (1978).

\bibitem{debate}A.Vilenkin, Phys. Rev. {\bf D58}, 067301 (1998),
gr-qc/9804051; gr-qc/9812027.

\bibitem{slih}A.O.Barvinsky and A.Yu.Kamenshchik, J. Cosmol.
Astropart. Phys. {\bf 09}, 014 (2006), hep-th/0605132.

\bibitem{lcb}A.O.Barvinsky and A.Yu.Kamenshchik, Phys. Rev. {\bf
D74}, 121502 (2006), hep-th/0611206.

\bibitem{why}A.O.Barvinsky, Phys. Rev. Lett. {\bf 99} (2007) 071301,
hep-th/0704.0083

\bibitem{BDK}A.O.Barvinsky, C.Deffayet
and A.Yu.Kamenshchik, JCAP {\bf 05} (2008) 020, arXiv:0801.2063.

\bibitem{AdS/CFT}J.Maldacena, Adv. Theor. Math. Phys. {\bf 2} 231 (1998);
Int. J. Theor.Phys. {\bf 38} 1113 (1999), hep-th/9711200; E.Witten,
Adv. Theor. Math. Phys. {\bf 2}, 253 (1998); S.S.Gubser,
I.R.Klebanov and A.M.Polyakov, Phys. Lett. {\bf B428}, 105 (1998),
hep-th/9802109.

\bibitem{Randall:1999vf}
  L.~Randall and R.~Sundrum,
  Phys.\ Rev.\ Lett.\  {\bf 83} (1999) 4690
  [arXiv:hep-th/9906064].

\bibitem{Gubser}S.Gubser, Phys. Rev. {\bf D63}, 084017 (2001),
hep-th/9912001.

\bibitem{HHR}S.W. Hawking, T. Hertog and H.S. Reall, Phys.Rev.
{\bf D62} (2000) 043501.

\bibitem{Page}D.N.Page, Phys. Rev. {\bf D 34}, 2267 (1986).

\bibitem{Starobinsky}A.A.Starobinsky, Phys. Lett. {\bf 91B}, 99 (1980).

\bibitem{Duffanomaly}M.J.Duff, Class. Quant. Grav {\bf 11}, 1387
(1994), hep-th/9308075.

\bibitem{Buchbinder}I.L.Buchbinder, Fortsch. Phys. {\bf 34}, 605
(1986).

\bibitem{Riegert}R.J.Riegert, Phys. Lett. {\bf 134 B}, 56 (1984);
P.O.Mazur and E.Mottola, Phys. Rev. {\bf D 64}, 104022 (2001).

\bibitem{FrTs}E.S.Fradkin and A.A.Tseytlin, Phys. Lett. {\bf 134 B}, 187
(1984).

\bibitem{BMZ}A.O.Barvinsky, A.G.Mirzabekian and V.V.Zhytnikov,
``Conformal decomposition of the effective action and covariant
curvature expansion", gr-qc/9510037.

\bibitem{Buchbinder2}I.L.Buchbinder, V.P.Gusynin and P.I.Fomin,
Sov. J. Nucl. Phys. {\bf 44}, 534 (1986); I.L.Buchbinder and
S.M.Kuzenko, Nucl. Phys. {\bf B274}, 653 (1986).

\bibitem{FHH}M.V.Fischetti, J.B.Hartle and B.L.Hu, Phys. Rev.
{\bf D 20}, 1757 (1979).

\bibitem{can}L.D.Faddeev, Theor. Math. Phys. {\bf 1}, 1 (1969).

\bibitem{BarvU}A.O.Barvinsky, Phys. Rep. {\bf 230}, 237 (1993);
Nucl. Phys. {\bf B 520}, 533 (1998).

\bibitem{geom}A.O.Barvinsky and V.Krykhtin,
Class. Quantum Grav. {\bf 10}, 1957 (1993); A.O.Barvinsky,
gr-qc/9612003; M.Henneaux and C.Teitelboim, {\em Quantization of
Gauge Sytems} (Princeton University Press, Princeton, 1992).

\bibitem{Star}A.A.Starobinsky, Gravit. Cosmol. {\bf 6}, 157 (2000),
astro-ph/9912054.

\bibitem{HSch}J.B. Hartle and K. Schleich, in {\em Quantum field
theory and quantum statistics}, 67 (eds. I.Batalin et al, Hilger,
Bristol, 1988); K. Schleich, Phys.Rev. {\bf D 36}, 2342 (1987).

\bibitem{BYork}D. Brown and J.W. York, Jr., Phys. Rev. {\bf D 47}, 1420
(1993), gr-qc/9209014.

\bibitem{WoodardTsamis}N.C.Tsamis and R.P.Woodard, Nucl. Phys. {\bf
B 474}, 235 (1996).

\bibitem{DvaliKhouryetal}G.Dvali, S.Hofmann and J.Khoury,
``Degravitation of the Cosmological Constant and Graviton Width",
hep-th/0703027.

\bibitem{dark}R.R. Caldwell, R. Dave and P.J. Steinhardt,
Phys. Rev. Lett. {\bf 80}, 1582-1585 (1998); L.-M. Wang, R.R.
Caldwell, J.P. Ostriker and P.J. Steinhardt, Astrophys. J. {\bf
530}, 17 (2000).

\bibitem{otherDE}A.Yu. Kamenshchik, U. Moschella and V. Pasquier,
Phys. Lett. {\bf B511}, 265 (2001); V. Sahni and A.A. Starobinsky,
Int. J. Mod. Phys. {\bf D9}, 373 (2000); {\bf D15}, 2105 (2006).

\bibitem{bigbrake}V. Gorini, A.Yu. Kamenshchik, U. Moschella and V. Pasquier,
Phys. Rev. {\bf D69}, 123512 (2004).

\bibitem{Dvalispecies}G.Dvali, ``Black Holes and Large N Species
Solution to the Hierarchy Problem'', arXiv:0706.2050; G.Dvali and
M.Redi, ``Black Hole Bound on the Number of Species and Quantum
Gravity at LHC", arXiv:0710.4344 [hep-th].

\bibitem{Polch}J.Polchinski, {\em String Theory} (Cambridge
University Press, Cambridge, 1998).



\bibitem{DuffLiu}M.J.Duff and J.T.Liu, Phys. Rev. Lett. {\bf 85}, 2052 (2000),
hep-th/0003237.

\bibitem{BinDefLan}P.Binetruy, C.Deffayet and D.Langlois, Phys.
Lett. {\bf 477} 275 (2000), hep-th/9910219.
  P.~Binetruy, C.~Deffayet, U.~Ellwanger and D.~Langlois,
  Phys.\ Lett.\  B {\bf 477} (2000) 285
  [arXiv:hep-th/9910219].


\bibitem{TseytlinLiu}Hong Liu and A.A. Tseytlin, Nucl.Phys. {\bf B533}, 88
(1998), hep-th/9804083.

\bibitem{DGP}
G.~R.~Dvali, G.~Gabadadze and M.~Porrati,
  Phys.\ Lett.\  B {\bf 485} (2000) 208
  [arXiv:hep-th/0005016].

\bibitem{bulkBH}
T.~Shiromizu, K.~i.~Maeda and M.~Sasaki,
  Phys.\ Rev.\  D {\bf 62}, 024012 (2000)
  [arXiv:gr-qc/9910076].
  P.~Kraus,
  JHEP {\bf 9912}, 011 (1999)
  [arXiv:hep-th/9910149].

\bibitem{DGPDeffayet}
C.Deffayet, Phys. Lett. {\bf B 502}, 199 (2001), hep-th/0010186.

\bibitem{DDG}
C.Deffayet, G.Dvali and G.Gabadadze, Phys. Rev. {\bf D 65}, 044023
(2002), astro-ph/0105068.


\bibitem{bigcentalcharge} N.Arkani-Hamed, S.Dimopoulos, G.Dvali
and G.Gabadadze, ``Non-Local Modification of Gravity and the
Cosmological Constant Problem'', hep-th/0209227.

\end{thebibliography}
\end{document}